\documentclass[12pt,preprint]{aastex}
\usepackage{multirow}
\usepackage{color}
\usepackage{rotating}
\usepackage{enumerate}
\usepackage[hyphens]{url}
\usepackage{hyperref}
\shorttitle{VERITAS Observations of Cygnus~X-3}
\shortauthors{Angelo Varlotta for the VERITAS Collaboration}

\begin{document}
\title{VERITAS Observations of the Microquasar Cygnus~X-3}

\author{
S.~Archambault\altaffilmark{1},
M.~Beilicke\altaffilmark{2},
W.~Benbow\altaffilmark{3},
K.~Berger\altaffilmark{4},
R.~Bird\altaffilmark{5},
A.~Bouvier\altaffilmark{6},
J.~H.~Buckley\altaffilmark{2},
V.~Bugaev\altaffilmark{2},
K.~Byrum\altaffilmark{7},
M.~Cerruti\altaffilmark{3},
X.~Chen\altaffilmark{8,9},
L.~Ciupik\altaffilmark{10},
M.~P.~Connolly\altaffilmark{11},
W.~Cui\altaffilmark{12,\dag},
C.~Duke\altaffilmark{13},
J.~Dumm\altaffilmark{14},
M.~Errando\altaffilmark{15},
A.~Falcone\altaffilmark{16},
S.~Federici\altaffilmark{8,9},
Q.~Feng\altaffilmark{12},
J.~P.~Finley\altaffilmark{12},
L.~Fortson\altaffilmark{14},
A.~Furniss\altaffilmark{6},
N.~Galante\altaffilmark{3},
G.~H.~Gillanders\altaffilmark{11},
S.~Griffin\altaffilmark{1},
S.~T.~Griffiths\altaffilmark{17},
J.~Grube\altaffilmark{10},
G.~Gyuk\altaffilmark{10},
D.~Hanna\altaffilmark{1},
J.~Holder\altaffilmark{4},
G.~Hughes\altaffilmark{9},
T.~B.~Humensky\altaffilmark{18},
P.~Kaaret\altaffilmark{17},
M.~Kertzman\altaffilmark{19},
Y.~Khassen\altaffilmark{5},
D.~Kieda\altaffilmark{20},
H.~Krawczynski\altaffilmark{2},
M.~J.~Lang\altaffilmark{11},
A.~S~Madhavan\altaffilmark{21},
G.~Maier\altaffilmark{9},
P.~Majumdar\altaffilmark{22,23},
S.~McArthur\altaffilmark{24},
A.~McCann\altaffilmark{25},
P.~Moriarty\altaffilmark{26},
R.~Mukherjee\altaffilmark{15},
D.~Nieto\altaffilmark{18},
A.~O'Faol\'{a}in de Bhr\'{o}ithe\altaffilmark{5},
R.~A.~Ong\altaffilmark{22},
A.~N.~Otte\altaffilmark{27},
D.~Pandel\altaffilmark{28},
N.~Park\altaffilmark{24},
J.~S.~Perkins\altaffilmark{29},
M.~Pohl\altaffilmark{8,9},
A.~Popkow\altaffilmark{22},
H.~Prokoph\altaffilmark{9},
J.~Quinn\altaffilmark{5},
K.~Ragan\altaffilmark{1},
J.~Rajotte\altaffilmark{1},
L.~C.~Reyes\altaffilmark{30},
P.~T.~Reynolds\altaffilmark{31},
G.~T.~Richards\altaffilmark{27},
E.~Roache\altaffilmark{3},
G.~H.~Sembroski\altaffilmark{12},
F.~Sheidaei\altaffilmark{20},
A.~W.~Smith\altaffilmark{20},
D.~Staszak\altaffilmark{1},
I.~Telezhinsky\altaffilmark{8,9},
M.~Theiling\altaffilmark{12},
J.~V.~Tucci\altaffilmark{12},
J.~Tyler\altaffilmark{1},
A.~Varlotta\altaffilmark{12,\dag},
S.~Vincent\altaffilmark{9},
S.~P.~Wakely\altaffilmark{24},
T.~C.~Weekes\altaffilmark{3},
A.~Weinstein\altaffilmark{21},
D.~A.~Williams\altaffilmark{6},
B.~Zitzer\altaffilmark{7}
(The VERITAS Collaboration)\\
\&\\
M.~L.~McCollough\altaffilmark{32}
(Smithsonian Astrophysical Observatory)
}

\altaffiltext{1}{Physics Department, McGill University, Montreal, QC H3A 2T8, Canada}
\altaffiltext{2}{Department of Physics, Washington University, St. Louis, MO 63130, USA}
\altaffiltext{3}{Fred Lawrence Whipple Observatory, Harvard-Smithsonian Center for Astrophysics, Amado, AZ 85645, USA}
\altaffiltext{4}{Department of Physics and Astronomy and the Bartol Research Institute, University of Delaware, Newark, DE 19716, USA}
\altaffiltext{5}{School of Physics, University College Dublin, Belfield, Dublin 4, Ireland}
\altaffiltext{6}{Santa Cruz Institute for Particle Physics and Department of Physics, University of California, Santa Cruz, CA 95064, USA}
\altaffiltext{7}{Argonne National Laboratory, 9700 S. Cass Avenue, Argonne, IL 60439, USA}
\altaffiltext{8}{Institute of Physics and Astronomy, University of Potsdam, 14476 Potsdam-Golm, Germany}
\altaffiltext{9}{DESY, Platanenallee 6, 15738 Zeuthen, Germany}
\altaffiltext{10}{Astronomy Department, Adler Planetarium and Astronomy Museum, Chicago, IL 60605, USA}
\altaffiltext{11}{School of Physics, National University of Ireland Galway, University Road, Galway, Ireland}
\altaffiltext{12}{Department of Physics, Purdue University, West Lafayette, IN 47907, USA; avarlott@purdue.edu, cui@purdue.edu}
\altaffiltext{13}{Department of Physics, Grinnell College, Grinnell, IA 50112-1690, USA}
\altaffiltext{14}{School of Physics and Astronomy, University of Minnesota, Minneapolis, MN 55455, USA}
\altaffiltext{15}{Department of Physics and Astronomy, Barnard College, Columbia University, NY 10027, USA}
\altaffiltext{16}{Department of Astronomy and Astrophysics, 525 Davey Lab, Pennsylvania State University, University Park, PA 16802, USA}
\altaffiltext{17}{Department of Physics and Astronomy, University of Iowa, Van Allen Hall, Iowa City, IA 52242, USA}

\altaffiltext{18}{Physics Department, Columbia University, New York, NY 10027, USA}
\altaffiltext{19}{Department of Physics and Astronomy, DePauw University, Greencastle, IN 46135-0037, USA}
\altaffiltext{20}{Department of Physics and Astronomy, University of Utah, Salt Lake City, UT 84112, USA}
\altaffiltext{21}{Department of Physics and Astronomy, Iowa State University, Ames, IA 50011, USA}
\altaffiltext{22}{Department of Physics and Astronomy, University of California, Los Angeles, CA 90095, USA}
\altaffiltext{23}{Saha Institute of Nuclear Physics, Kolkata 700064, India}
\altaffiltext{24}{Enrico Fermi Institute, University of Chicago, Chicago, IL 60637, USA}
\altaffiltext{25}{Kavli Institute for Cosmological Physics, University of Chicago, Chicago, IL 60637, USA}
\altaffiltext{26}{Department of Life and Physical Sciences, Galway-Mayo Institute of Technology, Dublin Road, Galway, Ireland}
\altaffiltext{27}{School of Physics and Center for Relativistic Astrophysics, Georgia Institute of Technology, 837 State Street NW, Atlanta, GA 30332-0430}
\altaffiltext{28}{Department of Physics, Grand Valley State University, Allendale, MI 49401, USA}
\altaffiltext{29}{N.A.S.A./Goddard Space-Flight Center, Code 661, Greenbelt, MD 20771, USA}
\altaffiltext{30}{Physics Department, California Polytechnic State University, San Luis Obispo, CA 94307, USA}
\altaffiltext{31}{Department of Applied Physics and Instrumentation, Cork Institute of Technology, Bishopstown, Cork, Ireland}
\altaffiltext{32}{Smithsonian Astrophysical Observatory, 60 Garden Street, Cambridge, MA 02138, USA}
\altaffiltext{\dag}{Authors to whom any correspondence should be addressed.}

\begin{abstract}

We report results from TeV gamma-ray observations of the microquasar Cygnus~X-3. The observations were made with the Very Energetic Radiation Imaging Telescope Array System (VERITAS) over a time period from 2007 June 11 to 2011 November 28. VERITAS is most sensitive to gamma rays at energies between 85~GeV to 30~TeV. The effective exposure time amounts to a total of about 44 hours, with the observations covering six distinct radio/X-ray states of the object. No significant TeV gamma-ray emission was detected in any of the states, nor with all observations combined. The lack of a positive signal, especially in the states where GeV gamma rays were detected, places constraints on TeV gamma-ray production in Cygnus X-3. We discuss the implications of the results.

\end{abstract}

\keywords{acceleration of particles - binaries: close - gamma rays: stars - X-rays: individual (Cygnus~X-3)}

\section{Introduction}\label{introduction}

Cygnus~X-3 was among the first X-ray sources to be discovered in the early days of X-ray astronomy. It lies in the Galactic plane, at a distance between 7~kpc and 10~kpc~\citep{Predehl-2000, Ling-2009}. It is a high-mass X-ray binary, with the companion star appearing to show the spectral characteristics of a Wolf-Rayet star~\citep{Van Kerkwijk-1996}. The nature of the compact object is still being debated. Cygnus~X-3 is known to produce intense radio flares, making it at times one of the brightest transient Galactic radio sources. The radio flares can last from a few days to several weeks. Relativistic jets have been seen during major flares ($S_{\nu}>$ 10 Jy at 15~GHz)~\citep{Mioduszewski-2001, Miller-Jones-2004}, with an inclination to the line of sight of $\lesssim$~14\arcdeg . This makes Cygnus~X-3 analogous to the extragalactic blazars, which constitute a major population of known TeV gamma-ray emitters. With an orbital period of only 4.8 hours, the compact object is thought to be enshrouded in the wind of the Wolf-Rayet star. 

Cygnus~X-3 has long been a prominent target for gamma-ray observations. Initially, there was contradictory evidence for gamma-ray emission at GeV energies: SAS-2 found a periodic signal~\citep{Lamb-1977}, while COS-B could not confirm it~\citep{Bennett-1977}. Then, numerous claims of detection of Cygnus~X-3 were made in the early days of ground-based gamma-ray experiments, spanning the TeV to PeV energy range (see discussions by \citealt{Weekes-1992} and \citealt{Ong-1998}). The claims were subsequently disputed by a critical analysis of the observations~\citep{Chardin-1989}. In subsequent years, Cygnus X-3 was observed with more sensitive ground-based instruments, including Whipple~\citep{O'Flaherty-1992}, CASA-MIA~\citep{Borione-1997}, HEGRA~\citep{Schilling-2001}, and MAGIC~\citep{Aleksic-2010}, but was not detected. At GeV energies, EGRET/CGRO found a gamma-ray source (2EG~J2033+4112) that is consistent with the position of Cygnus~X-3 (although the position error circle is quite large), but with no evidence for orbital modulation~\citep{Mori-1997}. The source has now been detected at GeV energies, with high confidence, independently with AGILE~\citep{Tavani-2009} and Fermi~LAT \citep{Abdo-2009}. Moreover, the orbital modulation of the gamma-ray emission has also been seen~\citep{Abdo-2009}.

In X-ray binaries, gamma rays may be produced by Compton upscattering of photons, either from the companion star or the accretion disk or both, by relativistic electrons accelerated in the jets of a stellar-mass black hole or in the shocked wind of a pulsar. Detailed models have been constructed for gamma-ray production and attenuation in the jets. In the case of Cygnus~X-3, the close proximity ($R_{d}\approx3\times10^{11}\,$cm), high temperature ($T_{*}\sim 10^{5}\,$K), and high luminosity ($L_{*}\sim{}10^{39}\,$erg s$^{-1}$) of the Wolf-Rayet star may result in the efficient Compton upscattering of stellar photons to produce gamma rays, as well as in the attenuation of the gamma rays via $\gamma\gamma$ pair production~\citep{Bednarek-2010}. Whether Cygnus~X-3 appears as a TeV gamma-ray emitter would depend on the competition between the production and attenuation processes. Theoretically, certain circumstances would favor TeV emission. These typically involve emitting regions at large perpendicular distances ($H\gtrsim{}10\,{R_{d}}$) from the orbital plane and orbital phases around the inferior conjunction~\citep{Bednarek-2010}. At lower (GeV) energies, the attenuation optical depth is much reduced, so the observed orbital modulation may be mainly associated with the production process~\citep{Dubus-2010,Bednarek-2010}. Alternatively, gamma rays may also be produced by the decay of $\pi^{0}$ particles, which result from the $p\,p$ collisions between the relativistic protons in the jets and the cold protons of the dense anisotropic stellar wind of the Wolf-Rayet star~\citep{Romero-2003}. 

Cygnus~X-3 is a persistent radio source. Its radio flux may vary by four orders of magnitude. Based on the long-term monitoring of the source with the Green Bank Interferometer (GBI), four radio states were identified~\citep{Waltman-1994}: quiescent state (60-140~mJy), minor flaring state ($\lesssim\,$1~Jy), quenched state ($\lesssim\,$30~mJy), and intermediate/major flaring state ($>$1~Jy). The major flaring state seems to follow the quenched state. The radio emission was subsequently found to be correlated with the hard X-ray emission~\citep{McCollough-1999}. The correlation is complex and varies with the state that the source is in: it is negative (anti-correlation) in the quiescent state but turns positive in the major flaring and quenched states. No apparent correlation has been observed in the minor flaring state. The radio emission is also correlated with the soft X-ray emission in certain states \citep{Hjalmarsdotter-2008}. This is expected because it is known that the soft and hard X-ray fluxes of Cygnus X-3 are generally (but not always) anti-correlated \citep{Choudhury-2002,Hjalmarsdotter-2008}. Based on the correlated radio/X-ray properties of the source, \citet{Szostek-2008} refined and expanded the definitions of the states. The new radio/X-ray states are now referred to as the \emph{quiescent}, \emph{minor-flaring}, \emph{suppressed}, \emph{quenched}, \emph{major-flaring} and \emph{post-flaring} states.

The AGILE and Fermi-LAT observations have shown that the gamma-ray emission from Cygnus X-3 is not steady but episodic. A careful examination of the gamma-ray activities of the source has revealed that gamma-ray production appears to be associated with transitions into or out of the radio \emph{quenched} state \citep{Koljonen-2010}. During a transition, the X-ray spectrum of the source becomes dominated by a soft X-ray component (with only a weak power-law component) as its radio flux goes down. For this reason, these time periods are now also referred to as the hypersoft state \citep{Koljonen-2010}. As such, the line between the hypersoft state and \emph{quenched} state is not always very clear in practice. Observationally, the hypersoft state is associated with major radio flares and, sometimes, the formation of jets~\citep{Koljonen-2010}. The latter might be the site of gamma-ray production. This signifies the importance of the hypersoft state to our understanding of Cygnus X-3 as a gamma-ray emitter. Unfortunately, the hypersoft state is very short in duration (lasting for $\lesssim$ 4-5 days), compared with other states, so it is often challenging to catch it with sensitive instruments of small field of view. 

In this work, we carried out a systematic search for gamma rays from Cygnus~X-3 at TeV energies with the Very Energetic Radiation Imaging Telescope Array System (VERITAS). The availability of the contemporaneous radio/X-ray observations of the source made it possible to extend the search to individual radio/X-ray states, particularly to the hypersoft state.

\section{VERITAS Observations}\label{TeV Gamma-Ray Observations}

VERITAS is a ground-based gamma-ray telescope array located at the Fred Lawrence Whipple Observatory in southern Arizona in the United States. It consists of four 12-meter imaging atmospheric Cherenkov telescopes, each with 499-pixel cameras, designed to detect the faint flashes of Cherenkov light from air showers initiated in the atmosphere by TeV gamma rays or cosmic rays. VERITAS can detect gamma rays in the energy range from 85~GeV to 30~TeV, with a maximum effective area of approximately 10$^5$ m$^2$. The energy resolution is about 15-25\%. At 1~TeV, the angular resolution is better than 0.1\arcdeg\ for an individual gamma event (68\% containment level). The pointing accuracy of VERITAS is $<$50\arcsec. VERITAS has a field of view (FoV) of about 3.5\arcdeg\ in angular diameter. In the summer of 2009, one of the four telescopes of the array was relocated to a different position, increasing the overall sensitivity of the array by about 30\%. After the relocation, VERITAS is capable of detecting sources at the flux level of 1\% of the Crab Nebula with a $\sim$25-hour exposure~\citep{Holder-2011}.

For this work, we used data from observations conducted between 2007 June 11 and 2011 November 28. The observations were conducted under varying weather and other conditions. The design of VERITAS also allows observations to be conducted under partial moonlight. To mitigate the night sky background fluctuations caused by the moonlight, the triggering threshold is increased in the camera photomultiplier tubes, which leads to a higher energy threshold. The triggering threshold in the photomultiplier tubes sets the first step in event discrimination in the telescope array. We carefully examined the data and included all of the observations which we believe can lead to reliable results. The total exposure time amounts to about 44 hours. More detailed information on the observations is shown in Table~\ref{table:Vdaily}.

The reduction of VERITAS data consists of multiple steps, including rejection of substandard data, flat fielding, pedestal subtraction,  pulse integration, image cleaning, parameterization of events, stereo reconstruction of shower direction and impact parameter, and gamma-ray/cosmic-ray separation. Briefly, the data from each participating telescope are first filtered for bad weather or issues with data acquisition, and are then charge integrated, pedestal subtracted and gain corrected. Each resulting image is cleaned and characterized to derive the moments of the light distributions~\citep{Hillas-1985}. The images of the same air shower from all participating telescopes are used to reconstruct the direction and impact parameter of the shower (see e.g.~\citealt{Krawczynski-2006}). This step requires characterizable images from three or more telescopes. In addition, to separate the gamma-ray events from the cosmic-ray events, we applied selection criteria (based on the energy and geometry of the events) to the events that survived the previous steps. The post-selection energy threshold is about 220~GeV at a 10\arcdeg\ zenith angle and 450~GeV at a 40\arcdeg\ zenith angle, which correspond approximately to the lowest and highest zenith angle of our data set, respectively. More details about VERITAS, the calibration procedure and the analysis technique can be found in~\citet{Acciari-2008}.

The VERITAS observation set was the product of different observation modes: it was composed of wobble-mode data on Cygnus~X-3, wobble-mode data taken on the TeV gamma-ray source TeV~J2032+4130~\citep{Aharonian-2005}, which is $\sim\,$30\arcmin\ from Cygnus~X-3, and data from tracking mode on the mid-point position between Cygnus~X-3 and TeV~J2032+4130. In wobble mode, the telescopes are pointed such that the source is always located at a fixed offset (0.5\arcdeg), alternately to the north, south, east and west of the camera center, for an unbiased estimation of the FoV background of the source region. In tracking mode, the telescopes were pointed alternatively to the east and west of the mid-point position between Cygnus~X-3 and TeV~J2032+4130. Due to the mixture of different observing modes, the data analysis used the ring background model~\citep{Berge-2007}. Briefly, the background estimate is derived for a trial source position from an annulus around the source region, which is dependent on the selection criteria. Due to the different offsets of the ring points with respect to the camera center as compared to the source position, a relative event rate, or \emph{acceptance}, correction needs to be applied to normalize the background rate. Any gamma-ray source in the FoV needs to be excluded from the background estimation as well. In our case, we excluded the pixels pointing at bright stars (with B magnitude less than 6) from the background regions. The nearby known TeV gamma-ray source, TeV~J2032+4130, was removed from subsequent analyses, to avoid incorrect estimation of the source and background rates of Cygnus~X-3 in the analysis. 

The data analysis on Cygnus~X-3 was performed with selection-criteria parameters based on the energy and geometry configuration of the gamma-ray initiated air showers, and modeled on the Crab Nebula. The selection criteria are optimized for a putative source with either a soft (6.6\% Crab at 200~GeV, spectral index: -4), medium (2\% Crab at 400~GeV, spectral index: -2.4) or hard (2\% Crab at~1~TeV, spectral index: -2.0) spectral index. The selection-criteria parameters tend to be looser for softer sources than for harder ones, to allow less event selection restrictions in the analysis. The acceptance correction was consequently generated over the whole data set for soft, medium and hard selection criteria and then applied to partial data sets (e.g., for individual states). For data taken with the initial VERITAS telescope array configuration (prior to August 2009) where telescope 1 (T1) and telescope 4 (T4) were in proximity to one another, all T1 and T4 simultaneous events were removed from analysis if no other telescope (T2 or T3) was triggered.

\section{Supporting Multi-Wavelength Observations}

To gain a broad-band perspective, we also examined data from observations at longer wavelengths. Particularly relevant to this work are contemporaneous gamma-ray observations of Cygnus X-3 at GeV energies made with the Large Area Telescope (LAT) aboard the \emph{Fermi Gamma-Ray Space Telescope} satellite~\citep{Atwood-2009}, as well as contemporaneous X-ray and radio observations, which make it possible to characterize the radio/X-ray states of the source.

\subsection{Fermi-LAT Observations}

In the default survey mode, the LAT scans the sky continuously and covers the whole sky once about every three hours. It  is sensitive to gamma rays in the nominal energy range of 0.02--300~GeV. Its on-axis effective area is between 8000 and 9000 cm$^2$ for energies $\gtrsim$1~GeV. The LAT has a very large FoV ($\sim$2.4 sr) and has an angular resolution of better than 0.1$^{\circ}$ at 1~GeV (for 68\% containment). For this work, we used the LAT data taken from 2008 August 5 to 2012 March 13.

The LAT data were processed with the Fermi Science Tools (v9r23p1), following the recommendations on event selection from the Fermi Science Support Center\footnote{\url{http://fermi.gsfc.nasa.gov/ssc/data/analysis/scitools/}}. Briefly, the events that have the highest probability of being gamma rays were selected by means of the Pass 7 V6 (P7\_V6) source class event selection cut with the {\tt gtselect} tool. In order to minimize contamination from Earth albedo photons, the time periods when Cygnus~X-3 was observed at zenith angles greater than 100$^{\circ}$ were eliminated from further analysis. The energy range was also limited from 0.1-100~GeV. 

For background modeling, we included all of the sources in the Fermi Large Area Telescope Second Source (2FGL) Catalog~\citep{Nolan-2012} that are in the vicinity of Cygnus~X-3. To account for possible intrinsic variability of the sources, we allowed the spectral parameters of the sources in a 5$^{\circ}$-radius region of interest (RoI) to vary in an unbinned maximum likelihood analysis. On the other hand, we froze the spectral parameters of the sources that are outside of the RoI but within a 22$^{\circ}$-radius source region at the 2FGL values. To minimize contamination from a bright nearby pulsar, PSR~J2032+4127 (about 30\arcsec\ away from Cygnus X-3), following \citet{Corbel-2012}, we excluded the times of its peak-pulse emission, based on the pulsar ephemeris\footnote{The pulsar ephemeris for PSR~J2032+4127 is available at: \href{http://www.slac.stanford.edu/~abdo/LATPulsarTimingModels/Latest/J2032+4127/J2032+4127_latest.par}{\\\texttt{http://www.slac.stanford.edu/\string~abdo/LATPulsarTimingModels/Latest/J2032+4127/J2032+4127\_latest.par}}} \citep{Ray-2011}. As for the Galactic and extragalactic diffuse gamma-ray backgrounds, we adopted the most recent models (\emph{gal\_2yearp7v6\_v0.fits} and \emph{iso\_p7v6source.txt}). We also modeled the emission from the Cygnus~Loop region with a template that is provided in the LAT Catalog Data Products. The instrument response function (IRF) used in this work is IRF P7SOURCE\_V6.

We derived, from background modeling, the best-fit spectral parameters of the sources in the RoI. We then fixed the parameters for all other sources, as well as the spectral index of Cygnus~X-3, and performed another unbinned maximum likelihood analysis, to produce a light curve of Cygnus~X-3 over the time period of interest. The statistical significance of each measurement is quantified by a maximum likelihood test statistic (TS; \citealt{Mattox-1996}), which corresponds roughly to $\sqrt{TS}$ $\sigma$ in Gaussian statistics.

\subsection{X-ray and Radio Observations}

Contemporaneous X-ray coverages of Cygnus~X-3 were provided by the All-sky Monitor (ASM) aboard the \emph{Rossi X-ray Timing Explorer} satellite~\citep{Levine-1996}, the Burst Alert Monitor (BAT) aboard the \emph{Swift} satellite~\citep{Barthelmy-2005}, and the Monitor of All-sky X-ray Image (MAXI) aboard the International Space Station~\citep{Matsuoka-2009}. The ASM and MAXI cover soft X-ray bands of 1.5--12 keV and 2--20 keV, respectively, while the BAT covers the hard X-ray band of 15--50 keV. For this work, we chose to use the ASM and MAXI data in a narrower (soft) band, to achieve a more accurate characterization of the states  \citep{Szostek-2008}. We weighted the measured count rates or fluxes of Cygnus X-3 (by $1/\sigma^2$), which are made publicly available by the instrument teams, and, if necessary, rebinned them to produce daily-averaged light curves. 

At radio wavelengths, Cygnus X-3 is monitored regularly with the \emph{Arcminute Microkelvin Imager-Large Array} (AMI-LA)\footnote{\protect{\url{http://www.mrao.cam.ac.uk/~guy/cx3/}}} at the Mullard Radio Astronomy Observatory in the UK. The AMI-LA consists of eight 12.8-meter Cassegrain antennas in a 2-dimensional array, with a baseline of $\sim\!120$ meters \citep{Zwart-2008}. It operates in six frequency bands covering the range of 13.9--18.2 GHz. Here, we used the data taken from 2008 May 26 to 2011 December 31. Note that no data were taken between 2006 June 19 and 2008 May 26, due to the major upgrade of the Ryle Radio Telescope to the AMI-LA. The weighted, daily-averaged light curve was used for this work.

\section{Results}\label{results}

\subsection{Blind Searches for TeV Gamma Rays} 

Using the full VERITAS data set, we found no significant ($> 5\sigma$) excess of TeV gamma rays from Cygnus~X-3 with the soft, medium and hard data cuts. The significance was calculated with the modified Eq.~(17) of \citet{Li-1983}, which is generalized for data sets with different source and background regions~\citep{Aharonian-2004}. The results are summarized in Table~\ref{table:states}. 

To derive a flux upper limit for each observing run, we calculated the total counts in the source region $N_{on}$, total counts in the background region $N_{off}$, and a scale factor $\alpha$, which is defined as the ratio of the areas of the (geometrical or parameter) regions from which source and background counts are derived. The scale factor may be different for different cuts. It may also vary from run to run, because, for instance, a bright star or known gamma source may need to be excluded from the background region in certain wobble configurations.
For the analyses of multiple data runs, individual $\alpha$'s were weighted by corresponding background counts and averaged to produce an effective $\alpha_{eff}$ for the runs. To account for varying zenith angle conditions, an average effective area $A_{eff}$ was constructed from individual effective areas for the runs. The flux upper limit was then derived from total $N_{on}$, total $N_{off}$, $\alpha_{eff}$, $A_{eff}$, and total effective exposure time, with the method of \citet{Rolke-2005}. 

Table~\ref{table:states} shows the 95\% confidence level (C.L.) integral flux upper limits derived with the full VERITAS data set. We chose as the lower limit for flux integration the energy threshold, which is defined as the energy at which the
differential rate of gamma-ray detection as a function of energy reaches its maximum. Different data cuts lead to different energy thresholds (also shown in the table). We should point out that we did not include systematic uncertainties in this or subsequent analyses. 

\subsubsection{Search for Episodic Emission}\label{episodic emission}

\begin{figure}[tp]
	\centering
		\includegraphics[width=1.0\textwidth]{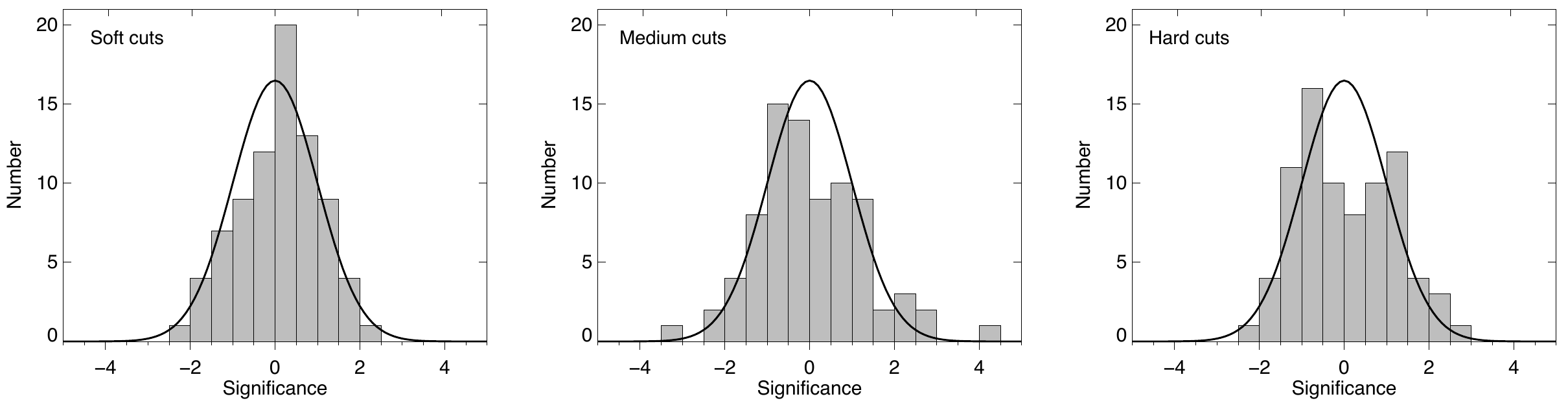}
	\caption{\footnotesize{Gaussian significance (in units of $\sigma$) distributions for VERITAS nightly searches. The results from different data cuts are shown separately. The Gaussian functions with mean zero and $\sigma$ one are shown in solid lines.}}
	\label{fig:signif-distr}
\end{figure}

We also conducted a blind search for episodic TeV gamma-ray emission from Cygnus~X-3. In this case, the VERITAS data runs were grouped on a night-by-night basis. As before, we selected events with the soft, medium and hard cuts, respectively, and followed the same procedure to reduce and analyze the data. Figure~\ref{fig:signif-distr} shows the distribution of the significance of excess for each set of cuts separately. The distributions are consistent with no significant TeV gamma-ray signal from Cygnus~X-3 (with the 99\% C.L. integral flux upper limits shown in the top panel of Figure 4 for individual nights).

\subsubsection{Search for Orbital Modulation}\label{orbital phase search}

\begin{figure}[t]
	\centering
		\includegraphics[width=0.72\textwidth]{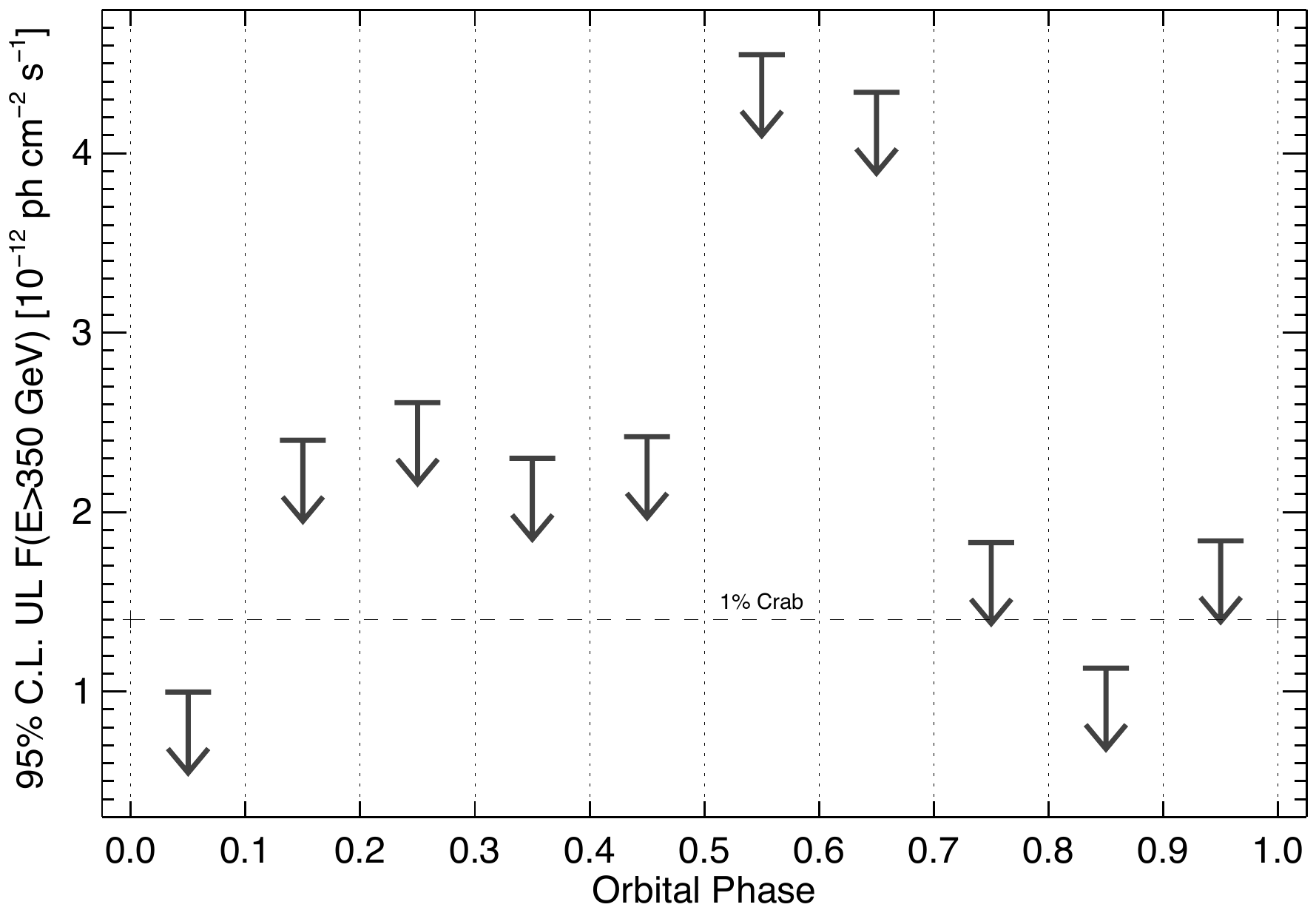}
	\caption{\footnotesize{VERITAS phase-folded 95\% C.L. integral ($E>350$~GeV) flux upper limits of Cygnus~X-3. For reference, the level of 1\% Crab is indicated (in dashed line). }}
	\label{fig:folded-lc}
\end{figure}

Considering that gamma-ray production could be concentrated in certain parts of the binary orbit, we folded the data from all observing runs into 10 phase bins, using the ephemeris of \citet{Zdziarski-2012c}. When a run spans multiple phase bins, we took care in dividing it so that the events fall in the correct bins. Again, we followed the same procedure to reduce and analyze the runs (or sub-runs) for each phase bin. We found no significant excess over the entire orbit. The 95\% C.L. integral flux upper limits (derived with the medium cuts) are shown in Figure~\ref{fig:folded-lc}.

\subsubsection{Spectral Constraints}\label{spectral limit search}

\begin{figure}[t]
	\centering
		\includegraphics[width=0.84\textwidth]{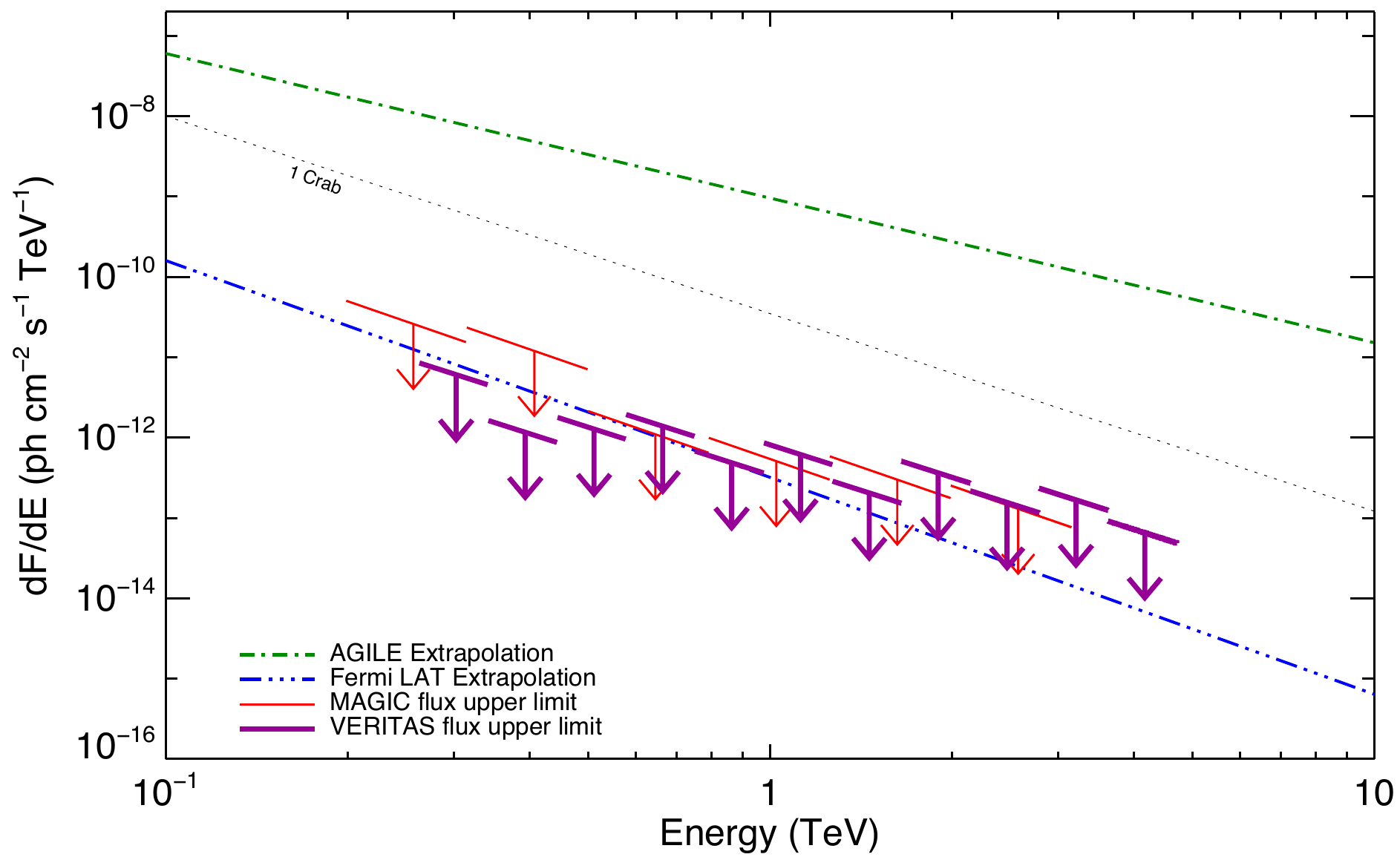}
	\caption{\footnotesize{VERITAS 95\% C.L. differential flux upper limits of Cygnus~X-3. For comparison, the published MAGIC upper limits are shown in (red) thin solid lines. See Figure 1 and Table 2 of~\citet{Aleksic-2010} for further details concerning the MAGIC results. A nominal spectrum of the Crab Nebula is shown in the (black) dotted line, for reference. The (blue) dot-dot-dot-dashed line and (green) dot-dashed line show the extrapolations of the power-law spectra measured with the Fermi LAT and AGILE at GeV energies, respectively.}}
	\label{fig:all-obs}
\end{figure}

To place constraints on the gamma-ray spectrum of Cygnus~X-3 at TeV energies, we also analyzed the data for selected energy ranges. The 95\% C.L. integral flux upper limits are given in Table~\ref{table:diff-spectrum}, and the corresponding differential flux upper limits shown in 
Figure~\ref{fig:all-obs}. We adopted logarithmic energy binning ($\Delta E/ E \sim 30\%$) for this analysis. The bins are coarser than the energy resolution of VERITAS but are sufficiently small to minimize any spectral dependence of the results. Such dependence may arise from the fact that the effective area is constructed, via Monte-Carlo simulations, with an assumed input spectrum (which, in this case, has a photon index of -2.4) and certain data cuts (which, in this case, are the medium cuts). Above about 5 TeV, the number of events that pass the cuts is so small that the results (not shown) become very uncertain. For comparison, we also plotted the published MAGIC differential flux upper limits \citep{Aleksic-2010} in the figure, as well as the extrapolation of the best-fit power-law spectra measured with AGILE and Fermi~LAT.  

\subsection{Targeted Searches for TeV Gamma Rays in Radio/X-ray States}

\begin{figure}[thbp]
	\centering
	\includegraphics[width=0.95\textwidth]{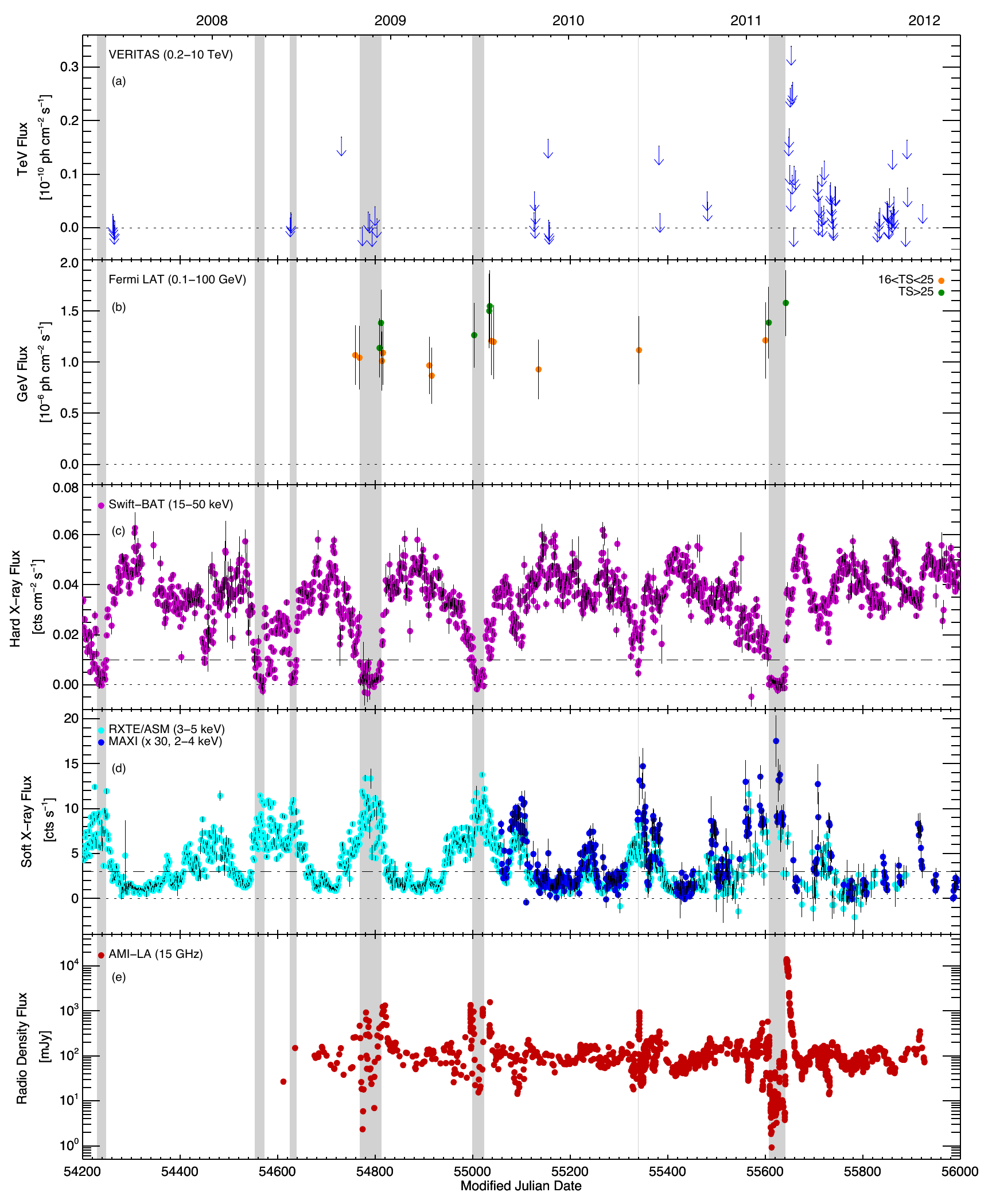}
	\caption{\footnotesize{Multi-wavelength light curves of Cygnus~X-3. Panels (a): TeV gamma ray. The VERITAS 99\% C.L. integral ($E>\,$263 GeV) flux upper limits are shown for individual nights.  
	(b): GeV gamma ray. The data points are color-coded by the detection significance: moderate significance (16$<\!TS\!<\!25$) in orange, and high significance ($TS\!>\!25$) in green. (c): Hard X-ray. (d): Soft X-ray. The ASM 3-5~keV measurements are shown in cyan and the MAXI 2-4~keV measurements in blue. Note that the MAXI flux values have been multiplied by 30 for clarity. (e): Radio. The AMI-LA 15~GHz measurements are shown. The shaded areas indicate the quenched state. The dot-dashed line in (c) and (d) shows roughly the threshold for transition into or out of the quenched state.}}
	\label{fig:mwplot}
\end{figure}

As mentioned in \S~\ref{introduction}, there appears to be evidence for gamma-ray production in Cygnus X-3 only in certain radio/X-ray states. For a more effective search, it is, therefore, important to characterize the states that the source is in. Fortunately, there were extensive X-ray and radio coverages of Cygnus X-3 that were contemporaneous with the VERITAS observations. We used the radio and (soft and hard) X-ray light curves of the source, as shown in Figure~\ref{fig:mwplot}, to distinguish the states, as defined in \citet{Szostek-2008}. We chose to divide the post-flaring state appropriately and merge it into the minor flaring and suppressed states. 

It is worth noting a few key features in the multi-wavelength light curves shown in Figure~\ref{fig:mwplot}. First, the anti-correlation between the soft and hard X-ray bands is apparent, comparing the ASM/MAXI and Swift BAT light curves. Second, the quenched state is not easily recognizable based on the radio light curve alone. It is, in fact, more apparent in the hard X-ray light curves, as it is when hard X-ray emission is quenched as well. To be more quantitative, we define the quenched state as the time when the Swift BAT flux goes below 0.01 cts cm$^{-2}$ s$^{-1}$ (or when the ASM flux goes above 3 cts s$^{-1}$, as the soft/hard X-ray anti-correlation suggests). Finally, the times of significant detections of Cygnus X-3 at GeV energies (see the Fermi-LAT light curve) do seem to align with the transitions into or out of the quenched state (i.e., the hypersoft state) quite well.

We grouped the VERITAS observing runs based on the radio/X-ray states, and carried out a search for TeV gamma rays from Cygnus~X-3 for each of the states. The analysis was made with the soft, medium, and hard cuts. The results are shown separately in Table~\ref{table:states}. No significant TeV gamma-ray signal was found in any of the searches.

\section{Discussion}\label{discussion}

The VERITAS observations of Cygnus X-3 covered the quenched state between 2008 October 30 and 2008 December 13 (MJD 54769--54813), when it was detected with AGILE (ATel \#1827,~\citealt{2008ATel.1827} and ATel \#1848,~\citealt{2008ATel.1848}). Unfortunately, there was only one VERITAS observation in the hypersoft/quenched state on 2008 December 4. The source was not detected at TeV energies. The derived flux upper limits are not very constraining (see Table~\ref{table:states}), due to limited VERITAS exposure. 

The VERITAS observations also covered the major flaring state (reaching a peak radio flux of $\sim$20~Jy) that followed the March 2011 quenched state. Due to the low elevation of the source and other observing constraints, VERITAS missed the peak of the radio flare (on 2011 March 24 or MJD 55644). The source was detected during this episode with the Fermi~LAT~\citep{Corbel-2012}. The highest LAT flux occurred on 2011 March 22, just before the peak of the radio flare. We failed to detect a signal from the source at TeV energies over the period from 2011 March 28 to April 5.

Based on the entire VERITAS data set, we derived, with the medium cuts, a 95\% C.L. integral flux upper limit of 0.7$\times10^{-12}$ ph cm$^{-2}$ s$^{-1}$ ($E>263$~GeV), which is about a factor of three lower than the published MAGIC value ($E>250$~GeV)~\citep{Aleksic-2010}. Note that the difference in energy thresholds between the two measurements amounts only to an effect of a few percent. We have also made a direct comparison of the VERITAS and MAGIC constraints on differential fluxes at various energies (see Figure~\ref{fig:all-obs}). Our upper limits are significantly lower than the MAGIC limits at lower energies. It should, however, be noted that we did not consider systematic uncertainties in our analyses, while the MAGIC results include a 30\% systematic uncertainty on flux. The VERITAS flux upper limits are compatible with the results of spectral modeling carried out by~\citet{Piano-2012}.

If we extrapolate the best-fit Fermi~LAT spectrum of Cygnus~X-3~\citep{Abdo-2009} to the VERITAS energy range, following a simple power law, we would expect an integral flux of $F(E>263$~GeV) = 1.8$\times10^{-12}$ ph cm$^{-2}$ s$^{-1}$, which is comparable to our 95\% C.L. flux upper limit. However, the uncertainties on the Fermi-LAT spectrum make it difficult to conclude that a spectral break or rollover would be required from GeV to TeV energies. The episodic nature of the gamma-ray emission from Cygnus~X-3 has made it even more difficult to compare Fermi~LAT and VERITAS measurements. This is illustrated by the fact that the published AGILE spectrum of Cygnus X-3 \citep{Tavani-2009} is higher and harder than the Fermi-LAT spectrum. If we extrapolate the best-fit power law to the AGILE spectrum into the VERITAS energy range, we would obtain an integral flux of $F(E>263$~GeV) = 3.5$\times10^{-9}$ ph cm$^{-2}$ s$^{-1}$ (with large uncertainties). More sophisticated spectral modeling is required to connect the Fermi-LAT and VERITAS data more physically (e.g., \citealt{Piano-2012}) but it is beyond the scope of this work.

For microquasars, the kinetic power of the jets is of the order of $\sim\!10^{38}$ erg s$^{-1}$, which is comparable to the bolometric luminosity of Cygnus~X-3 (assuming a distance of 9~kpc; ~\citealt{Hjalmarsdotter-2008}). Our flux upper limit corresponds to an upper limit on the TeV gamma-ray luminosity $\approx 6\times10^{33}$ erg s$^{-1}$. This implies a maximum gamma-ray conversion efficiency of the order $10^{-4}$-$10^{-5}$. In the context of leptonic models, \citet{Bednarek-2010} predicted a gamma-ray luminosity of $\approx{}10^{32}$ erg s$^{-1}$ for Cygnus~X-3, assuming steady emission. This is discouraging for the current generation of ground-based gamma-ray facilities, although the episodic nature of GeV gamma-ray emission from the source argues for more patience. In our case, the most interesting radio/X-ray state (i.e., the hypersoft state) has hardly been covered (see Table~\ref{table:states}). A concerted, multi-wavelength effort to target this state will likely be a more effective (and resource conserving) strategy for moving forward.

\section{Acknowledgments}
\label{sec:acknowledgments}

This work has made use of high-level data products provided by the ASM/RXTE, MAXI, and BAT/Swift teams. The AMI-LA radio results were obtained from a public archive maintained by Guy Pooley. Paul Ray is thanked for making available the ephemeris for PSR J2032+4127. A.V. and W.C. gratefully acknowledge financial support from NASA through a Fermi GI grant (NNX11AP90G) and from Purdue University. The VERITAS operation is supported by grants from the U.S. Department of Energy Office of Science, the U.S. National Science Foundation and the Smithsonian Institution, by NSERC in Canada, by Science Foundation Ireland (SFI 10/RFP/AST2748) and by STFC in the U.K. We acknowledge the excellent work of the technical support staff at the Fred Lawrence Whipple Observatory and at the collaborating institutions in the construction and operation of the instrument. 

{\it Facilities:} \facility{VERITAS}, \facility{Swift}, \facility{RXTE}, \facility{MAXI}, \facility{Fermi LAT}, \facility{AGILE}, \facility{AMI-LA}

\clearpage

\begin{deluxetable}{ccccccc}
\tabletypesize{\scriptsize}
\centering
\tablecolumns{2}
\tablewidth{0pt}
\tablecaption{Summary of VERITAS Observations
\label{table:Vdaily}}
\tablehead{
\colhead{MJD} &
\colhead{Calendar} &
\colhead{X-ray} &
\colhead{Observing Time} &
\colhead{Elevation} &
\colhead{N$_{\mathrm{tel}}$} \\
\colhead{} &
\colhead{Date} &
\colhead{State} &
\colhead{(min)} &
\colhead{Range} &
\colhead{}
}
\startdata
54262 & 2007/06/11 & Minor flaring & 20 & $65^{\mathrm{o}}-69^{\mathrm{o}}$ & 3 \\ 
54263 & 2007/06/12 & Minor flaring & 40 & $70^{\mathrm{o}}-77^{\mathrm{o}}$ & 3 \\ 
54264 & 2007/06/13 & Minor flaring & 119.5 & $62^{\mathrm{o}}-80^{\mathrm{o}}$ & 3 \\ 
54265 & 2007/06/14 & Minor flaring & 80 & $72^{\mathrm{o}}-80^{\mathrm{o}}$ & 3 \\ 
54266 & 2007/06/15 & Minor flaring & 40 & $72^{\mathrm{o}}-78^{\mathrm{o}}$ & 3 \\ 
54626 & 2008/06/09 & Quenched & 40 & $76^{\mathrm{o}}-80^{\mathrm{o}}$ & 4 \\ 
54627 & 2008/06/10 & Quenched & 40 & $76^{\mathrm{o}}-80^{\mathrm{o}}$ & 4 \\ 
54628 & 2008/06/11 & Quenched & 20 & $80^{\mathrm{o}}-81^{\mathrm{o}}$ & 4 \\ 
54731 & 2008/09/22 & Suppressed & 20 & $76^{\mathrm{o}}-78^{\mathrm{o}}$ & 4 \\ 
54774 & 2008/11/04 & Quenched & 20 & $59^{\mathrm{o}}-63^{\mathrm{o}}$ & 4 \\ 
54786 & 2008/11/16 & Major flaring & 60 & $59^{\mathrm{o}}-72^{\mathrm{o}}$ & 4 \\ 
54789 & 2008/11/19 & Quenched & 60 & $64^{\mathrm{o}}-68^{\mathrm{o}}$ & 4 \\ 
54794 & 2008/11/24 & Quenched & 40 & $54^{\mathrm{o}}-60^{\mathrm{o}}$ & 4 \\ 
54800 & 2008/11/30 & Quenched & 20 & $54^{\mathrm{o}}-58^{\mathrm{o}}$ & 3 \\ 
54804 & 2008/12/04 & Hypersoft\tablenotemark{a} & 20 & $53^{\mathrm{o}}-56^{\mathrm{o}}$ & 4 \\
55126 & 2009/10/22 & Quiescent & 20 & $74^{\mathrm{o}}-76^{\mathrm{o}}$ & 4 \\ 
55127 & 2009/10/23 & Quiescent & 36 & $67^{\mathrm{o}}-74^{\mathrm{o}}$ & 4 \\ 
55128 & 2009/10/24 & Quiescent & 77 & $59^{\mathrm{o}}-79^{\mathrm{o}}$ & 4 \\ 
55129 & 2009/10/25 & Quiescent & 40 & $65^{\mathrm{o}}-74^{\mathrm{o}}$ & 4 \\ 
55155 & 2009/11/20 & Quiescent & 20 & $59^{\mathrm{o}}-62^{\mathrm{o}}$ & 3 \\ 
55156 & 2009/11/21 & Quiescent & 40 & $56^{\mathrm{o}}-64^{\mathrm{o}}$ & 4 \\ 
55157 & 2009/11/22 & Quiescent & 20 & $56^{\mathrm{o}}-59^{\mathrm{o}}$ & 4 \\ 
55158 & 2009/11/23 & Quiescent & 16 & $64^{\mathrm{o}}-67^{\mathrm{o}}$ & 4 \\ 
55382 & 2010/07/05 & Minor flaring & 20 & $72^{\mathrm{o}}-76^{\mathrm{o}}$ & 4 \\ 
55384 & 2010/07/07 & Minor flaring & 4 & $80^{\mathrm{o}}-80^{\mathrm{o}}$ & 4 \\ 
55481 & 2010/10/12 & Quiescent & 40 & $75^{\mathrm{o}}-80^{\mathrm{o}}$ & 4 \\ 
55482 & 2010/10/13 & Quiescent & 40 & $69^{\mathrm{o}}-77^{\mathrm{o}}$ & 4 \\ 
55648 & 2011/03/28 & Major flaring & 20 & $42^{\mathrm{o}}-46^{\mathrm{o}}$ & 4 \\ 
55649 & 2011/03/29 & Major flaring & 20 & $42^{\mathrm{o}}-46^{\mathrm{o}}$ & 3 \\ 
55650 & 2011/03/30 & Major flaring & 20 & $43^{\mathrm{o}}-48^{\mathrm{o}}$ & 4 \\ 
55651 & 2011/03/31 & Major flaring & 28 & $45^{\mathrm{o}}-51^{\mathrm{o}}$ & 3 \\ 
55652 & 2011/04/01 & Major flaring & 20 & $42^{\mathrm{o}}-46^{\mathrm{o}}$ & 3 \\ 
55653 & 2011/04/02 & Major flaring & 20 & $42^{\mathrm{o}}-46^{\mathrm{o}}$ & 4 \\ 
55654 & 2011/04/03 & Major flaring & 15 & $48^{\mathrm{o}}-50^{\mathrm{o}}$ & 4 \\ 
55655 & 2011/04/04 & Major flaring & 20 & $48^{\mathrm{o}}-50^{\mathrm{o}}$ & 4 \\ 
55656 & 2011/04/05 & Major flaring & 20 & $48^{\mathrm{o}}-52^{\mathrm{o}}$ & 4 \\ 
55658 & 2011/04/07 & Minor flaring & 8 & $52^{\mathrm{o}}-53^{\mathrm{o}}$ & 3 \\ 
55659 & 2011/04/08 & Minor flaring & 23 & $50^{\mathrm{o}}-53^{\mathrm{o}}$ & 4 \\ 
55662 & 2011/04/11 & Minor flaring & 6 & $55^{\mathrm{o}}-56^{\mathrm{o}}$ & 3 \\ 
55707 & 2011/05/26 & Minor flaring & 20 & $72^{\mathrm{o}}-74^{\mathrm{o}}$ & 4 \\ 
55708 & 2011/05/27 & Minor flaring & 20 & $60^{\mathrm{o}}-64^{\mathrm{o}}$ & 4 \\ 
55709 & 2011/05/28 & Minor flaring & 96 & $65^{\mathrm{o}}-80^{\mathrm{o}}$ & 4 \\ 
55710 & 2011/05/29 & Minor flaring & 52 & $70^{\mathrm{o}}-77^{\mathrm{o}}$ & 3/4\tablenotemark{b} \\ 
55713 & 2011/06/01 & Minor flaring & 20 & $59^{\mathrm{o}}-62^{\mathrm{o}}$ & 4 \\ 
55715 & 2011/06/03 & Minor flaring & 40 & $77^{\mathrm{o}}-80^{\mathrm{o}}$ & 4 \\ 
55716 & 2011/06/04 & Minor flaring & 20 & $70^{\mathrm{o}}-74^{\mathrm{o}}$ & 4 \\ 
55717 & 2011/06/05 & Minor flaring & 20 & $79^{\mathrm{o}}-81^{\mathrm{o}}$ & 4 \\ 
55720 & 2011/06/08 & Minor flaring & 20 & $76^{\mathrm{o}}-78^{\mathrm{o}}$ & 4 \\ 
55721 & 2011/06/09 & Minor flaring & 10 & $74^{\mathrm{o}}-75^{\mathrm{o}}$ & 4 \\ 
55733 & 2011/06/21 & Minor flaring & 14 & $45^{\mathrm{o}}-48^{\mathrm{o}}$ & 4 \\ 
55734 & 2011/06/22 & Minor flaring & 10 & $64^{\mathrm{o}}-68^{\mathrm{o}}$ & 4 \\ 
55735 & 2011/06/23 & Minor flaring & 46 & $59^{\mathrm{o}}-69^{\mathrm{o}}$ & 4 \\ 
55736 & 2011/06/24 & Minor flaring & 84 & $74^{\mathrm{o}}-80^{\mathrm{o}}$ & 4 \\ 
55737 & 2011/06/25 & Minor flaring & 40 & $69^{\mathrm{o}}-73^{\mathrm{o}}$ & 4 \\ 
55738 & 2011/06/26 & Quiescent & 59.5 & $74^{\mathrm{o}}-80^{\mathrm{o}}$ & 4 \\ 
55739 & 2011/06/27 & Quiescent & 95 & $71^{\mathrm{o}}-80^{\mathrm{o}}$ & 4 \\ 
55740 & 2011/06/28 & Quiescent & 30 & $76^{\mathrm{o}}-80^{\mathrm{o}}$ & 3 \\ 
55743 & 2011/07/01 & Quiescent & 20 & $76^{\mathrm{o}}-80^{\mathrm{o}}$ & 4 \\ 
55744 & 2011/07/02 & Quiescent & 20 & $76^{\mathrm{o}}-78^{\mathrm{o}}$ & 4 \\ 
55830 & 2011/09/26 & Quiescent & 80 & $78^{\mathrm{o}}-80^{\mathrm{o}}$ & 3/4\tablenotemark{b} \\ 
55833 & 2011/09/29 & Quiescent & 48 & $74^{\mathrm{o}}-80^{\mathrm{o}}$ & 4 \\ 
55834 & 2011/09/30 & Quiescent & 52 & $64^{\mathrm{o}}-79^{\mathrm{o}}$ & 3/4\tablenotemark{b} \\ 
55835 & 2011/10/01 & Quiescent & 43.5 & $66^{\mathrm{o}}-75^{\mathrm{o}}$ & 4 \\ 
55850 & 2011/10/16 & Quiescent & 28 & $74^{\mathrm{o}}-80^{\mathrm{o}}$ & 4 \\ 
55851 & 2011/10/17 & Quiescent & 80 & $72^{\mathrm{o}}-80^{\mathrm{o}}$ & 4 \\ 
55852 & 2011/10/18 & Quiescent & 71 & $68^{\mathrm{o}}-80^{\mathrm{o}}$ & 4 \\ 
55853 & 2011/10/19 & Quiescent & 51 & $70^{\mathrm{o}}-79^{\mathrm{o}}$ & 4 \\ 
55854 & 2011/10/20 & Quiescent & 111 & $70^{\mathrm{o}}-79^{\mathrm{o}}$ & 4 \\ 
55855 & 2011/10/21 & Quiescent & 20 & $58^{\mathrm{o}}-60^{\mathrm{o}}$ & 4 \\ 
55858 & 2011/10/24 & Quiescent & 20 & $59^{\mathrm{o}}-61^{\mathrm{o}}$ & 4 \\ 
55860 & 2011/10/26 & Quiescent & 20 & $58^{\mathrm{o}}-61^{\mathrm{o}}$ & 4 \\ 
55861 & 2011/10/27 & Quiescent & 17 & $70^{\mathrm{o}}-74^{\mathrm{o}}$ & 4 \\ 
55862 & 2011/10/28 & Quiescent & 72 & $59^{\mathrm{o}}-80^{\mathrm{o}}$ & 4 \\ 
55863 & 2011/10/29 & Quiescent & 35 & $74^{\mathrm{o}}-80^{\mathrm{o}}$ & 4 \\ 
55864 & 2011/10/30 & Quiescent & 15 & $76^{\mathrm{o}}-80^{\mathrm{o}}$ & 4 \\ 
55865 & 2011/10/31 & Quiescent & 40 & $66^{\mathrm{o}}-76^{\mathrm{o}}$ & 4 \\ 
55888 & 2011/11/23 & Quiescent & 36 & $60^{\mathrm{o}}-68^{\mathrm{o}}$ & 4 \\ 
55891 & 2011/11/26 & Quiescent & 20 & $61^{\mathrm{o}}-64^{\mathrm{o}}$ & 4 \\ 
55892 & 2011/11/27 & Quiescent & 40 & $56^{\mathrm{o}}-64^{\mathrm{o}}$ & 4 \\ 
55893 & 2011/11/28 & Quiescent & 20 & $52^{\mathrm{o}}-56^{\mathrm{o}}$ & 4 \\ 
\enddata
\tablenotetext{a}{The hypersoft state consists of the data run contained within the quenched state. See text.}
\tablenotetext{b}{One telescope was taken out of the operation during the run.}
\tablecomments{The column N$_{\mathrm{tel}}$ shows the number of working telescopes.}
\end{deluxetable}

\begin{deluxetable}{cccccccccc}
\tabletypesize{\scriptsize}
\centering
\tablecolumns{10}
\tablewidth{0pt}
\tablecaption{Results from Gamma-Ray Searches
\label{table:states}}
\tablehead{
\colhead{Spectral} &
\colhead{Exposure} &
\colhead{Elevation} &
\colhead{On} &
\colhead{Off} &
\colhead{$\alpha_{eff}$} &
\colhead{Excess} &
\colhead{Significance} &
\colhead{Energy} &
\multicolumn{1}{c}{Flux Upper Limit} \\
\colhead{State} &
\colhead{Time} &
\colhead{Range} &
\colhead{Events} &
\colhead{Events} &
\colhead{} &
\colhead{Events} &
\colhead{($\sigma$)} &
\colhead{Threshold} &
\colhead{($10^{-12}$cm$^{-2}$s$^{-1})$} \\
\colhead{} &
\colhead{(hours)} &
\colhead{} &
\colhead{$N_{on}$} &
\colhead{$N_{off}$} &
\colhead{} &
\colhead{$N_{ex}$} &
\colhead{} &
\colhead{(GeV)} &
\colhead{}
}
\startdata
\multicolumn{10}{c}{Soft Cuts} \\ [0.5ex]
All & 44.70 & $42^{\mathrm{o}}-81^{\mathrm{o}}$ & 17509 & 125799 & 0.136 & 400.3 & 0.6 & 182 & 5.0 \\
\hline
Quiescent & 23.04 & $52^{\mathrm{o}}-80^{\mathrm{o}}$ & 9046 & 65596 & 0.136 & 124.9 & 0.3 & 182 & 4.6 \\
Minor flaring & 13.68 & $45^{\mathrm{o}}-81^{\mathrm{o}}$ & 4032 & 28865 & 0.138 & 48.6 & 0.6 & 200 & 6.1 \\
Suppressed & 0.30 & $76^{\mathrm{o}}-78^{\mathrm{o}}$ & 162 & 1069 & 0.156 & -4.8 & -0.4 & 200 & 64.6 \\
Quenched & 4.24 & $54^{\mathrm{o}}-81^{\mathrm{o}}$ & 2410 & 16923 & 0.142 & 6.9 & 0.1 & 200 & 20.5 \\
Hypersoft\tablenotemark{a} & 0.30 & $53^{\mathrm{o}}-63^{\mathrm{o}}$ & 180 & 1360 & 0.142 & -13.1 & -0.9 & 316 & 29.0 \\
Major flaring & 3.44 & $42^{\mathrm{o}}-72^{\mathrm{o}}$ & 1859 & 13344 & 0.138 & 17.5 & 0.4 & 316 & 10.9 \\
\hline
\multicolumn{10}{c}{Medium Cuts} \\ [0.5ex]
All & 44.70 & $42^{\mathrm{o}}-81^{\mathrm{o}}$ & 1200 & 26176 & 0.046 & -4.1 & -0.1 & 263 & 0.7 \\
\hline
Quiescent & 23.04 & $52^{\mathrm{o}}-80^{\mathrm{o}}$ & 654 & 15268 & 0.046 & -48.3 & -1.7 & 263 & 0.5 \\
Minor flaring & 13.68 & $45^{\mathrm{o}}-81^{\mathrm{o}}$ & 343 & 6813 & 0.046 & 29.6 & 1.5 & 288 & 2.1 \\
Suppressed & 0.30 & $76^{\mathrm{o}}-78^{\mathrm{o}}$ & 11 & 94 & 0.047 & 6.6 & 2.5 & 288 & 41.8 \\
Quenched & 4.24 & $54^{\mathrm{o}}-81^{\mathrm{o}}$ & 96 & 2097 & 0.047 & -2.6 & -0.1 & 347 & 2.5 \\
Hypersoft\tablenotemark{a} & 0.30 & $53^{\mathrm{o}}-63^{\mathrm{o}}$ & 8 & 205 & 0.045 & -1.2 & -0.4 & 457 & 9.0 \\
Major flaring & 3.44 & $42^{\mathrm{o}}-72^{\mathrm{o}}$ & 96 & 1904 & 0.047 & 6.5 & 0.7 & 550 & 2.2 \\
\hline
\multicolumn{10}{c}{Hard Cuts} \\ [0.5ex]
All & 44.70 & $42^{\mathrm{o}}-81^{\mathrm{o}}$ & 145 & 3305 & 0.045 & -3.7 & -0.2 & 603 & 0.2 \\
\hline
Quiescent & 23.04 & $52^{\mathrm{o}}-80^{\mathrm{o}}$ & 68 & 1936 & 0.045 & -19.1 & -2.0 & 603 & 0.1 \\
Minor flaring & 13.68 & $45^{\mathrm{o}}-81^{\mathrm{o}}$ & 46 & 831 & 0.046 & 7.8 & 1.2 & 603 & 0.6 \\
Suppressed & 0.30 & $76^{\mathrm{o}}-78^{\mathrm{o}}$ & 1 & 14 & 0.045 & 0.4 & 0.4 & 603 & 10.2 \\
Quenched & 4.24 & $54^{\mathrm{o}}-81^{\mathrm{o}}$ & 13 & 281 & 0.046 & 0.1 & 0.0 & 871 & 0.9 \\
Hypersoft\tablenotemark{a} & 0.30 & $53^{\mathrm{o}}-63^{\mathrm{o}}$ & 3 & 25 & 0.042 & 2.0 & 1.5 & 871 & 9.2 \\
Major flaring & 3.44 & $42^{\mathrm{o}}-72^{\mathrm{o}}$ & 17 & 244 & 0.047 & 5.5 & 1.5 & 955 & 1.2 \\
\enddata
\tablenotetext{a}{The hypersoft state consists of the data run from 2008/12/04 (MJD 54804) and is a data run subset contained within the quenched state.}
\tablecomments{Flux upper limits are given at the 95\% C.L, and for each row are calculated from the energy threshold. The column $\alpha_{eff}$ shows the effective scale factor for the background calculation (see \S~\ref{results}). }
\end{deluxetable}

\begin{deluxetable}{ccccccc}
\tabletypesize{\scriptsize}
\centering
\tablecolumns{2}
\tablewidth{0pt}
\tablecaption{Flux Upper Limits for Selected Energy Ranges
\label{table:diff-spectrum}}
\tablehead{
\colhead{Energy Range} &
\colhead{On} &
\colhead{Off} &
\colhead{$\alpha_{eff}$} &
\colhead{Excess} &
\colhead{Significance} &
\colhead{Flux Upper Limit} \\
\colhead{(TeV)} &
\colhead{Events} &
\colhead{Events} &
\colhead{} &
\colhead{Events} &
\colhead{($\sigma$)} &
\colhead{($10^{-12}$cm$^{-2}$s$^{-1})$} \\
\colhead{} &
\colhead{$N_{on}$} &
\colhead{$N_{off}$} &
\colhead{} &
\colhead{$N_{ex}$} &
\colhead{}
}
\startdata
0.263-0.342 & 230 & 4726 & 0.046 & 12.6 & 0.8 & 0.5 \\
0.342-0.445 & 151 & 3801 & 0.046 & -23.8 & -1.9 & 0.1 \\
0.445-0.578 & 126 & 2905 & 0.046 & -7.6 & -0.7 & 0.2 \\
0.578-0.751 & 102 & 2229 & 0.046 & -0.5 & -0.1 & 0.2 \\
0.751-0.977 & 65 & 1663 & 0.046 & -11.5 & -1.3 & 0.1 \\
0.977-1.269 & 58 & 1253 & 0.046 & 0.4 & 0.0 & 0.2 \\
1.269-1.650 & 36 & 1033 & 0.046 & -11.5 & -1.7 & 0.1 \\
1.650-2.145 & 39 & 795 & 0.046 & 2.4 & 0.4 & 0.2 \\
2.145-2.789 & 25 & 627 & 0.046 & -3.8 & -0.7 & 0.1 \\
2.789-3.626 & 20 & 447 & 0.046 & -0.6 & -0.1 & 0.1 \\
3.626-4.713 & 14 & 354 & 0.046 & -2.3 & -0.6 & 0.1 \\

\enddata
\tablecomments{As for Table~\ref{table:states}, but for selected energy ranges. }
\end{deluxetable}

\clearpage
\end{document}